\documentstyle[preprint,aps,epsf,prb]{revtex}
\begin{document}
\draft
\title{Interband effects in the $c$-axis optical conductivity in
YBa$_2$Cu$_3$O$_{7-\delta}$}
\author{W. A. Atkinson\cite{bill} and J. P. Carbotte}
\address{Department of Physics and Astronomy, McMaster University,
Hamilton, Ontario, Canada L8S 4M1}
\date{\today}

\maketitle
\begin{abstract}
The normal state optical conductivity is calculated for a layered metal 
with two layers per unit cell coupled through a transverse hopping matrix 
element $t_\perp$.  The optical response involves an interband term in 
addition to the more familiar intraband term which leads to 
the usual Drude form.  The interband term is only weakly temperature 
dependent, even for an inelastic scattering rate which is linear in T.  
It gives a $c$-axis response which extends in frequency over the entire 
band width although there can be structure on this energy scale which 
reflects details of the electronic structure.  In particular, at low energy, 
the $c$-axis response can develop a gap or pseudogap as the temperature is 
lowered.  At high temperature, a Drude response will be seen only if the 
intraband transitions, which are of order $t_\perp^4$, become important 
compared with the interband transitions which are of order $t_\perp^2$.
\end{abstract}
%

\narrowtext
\section{Introduction}
Models of the high $T_c$ oxides often start with a single isolated CuO$_2$ 
layer.  Other structural elements within the unit cell are usually labelled
as charge reservoirs or barrier layers and are ignored.  
Yet it is precisely the nature of the coupling between the different layers
that determines $c$-axis transport properties, which are 
found to display a rich variety of behaviours and are often 
anomalous.\cite{Cooper,Iye} The anomalous nature of the $c$-axis 
conductivity has been variously interpreted as suggesting that
interlayer coupling is an essential piece of the superconducting mechanism
\cite{Wheatley}, that the interlayer coupling is incoherent due to
impurity or phononic scattering\cite{Graf,Zha,Rojo,Kumar} or thermal 
fluctuations
\cite{Leggett}, or that the CuO$_2$ layers are in a non-Fermi liquid
state\cite{Anderson}.  In this work, we suggest that the $c$-axis 
optical conductivity in YBa$_2$Cu$_3$O$_{7-\delta}$ (YBCO$_{7-\delta}$)
can be explained by proper consideration of the multilayered structure of the
unit cell.

In considering $c$-axis properties, it is important to distinguish between 
the coupling between various conducting layers within a unit cell (which 
can contain several planes) and the intercell coupling which could involve 
some barrier layer.  It is this latter coupling that probably governs the 
size of the anisotropy observed between $c$-axis\cite{Cooper,Iye,Friedmann,%
Homes,Homes2} and $ab$-plane properties.  On the other hand, 
the large in-plane anisotropy between $a$- and $b$-directions\cite{Bonn,%
Tallon,Basov,Gagnon,Zhang} (along the CuO chains) observed in YBCO
is related more closely to the properties within a unit cell.  The 
actual situation is clearly quite complex.  For example, the unit cell in 
YBCO consists of a bilayer of two CuO$_2$ planes separated by a CuO chain 
layer.  In addition, the chains are only completed in 
YBa$_2$Cu$_3$O$_7$ and the effect of oxygen doping on the chain Fermi 
surface is not well understood nor is the partition of holes between 
planes and chains.  Because
of these uncertainties, it is necessary, at this stage, to use a simplified 
model and to set specific but limited aims.

Having recognized that several distinct transverse hopping matrix elements 
come into a complete description of the $c$-axis properties of the oxides, 
we will, nevertheless, limit ourselves here to a model of two layers per 
unit cell coupled through a single transverse matrix element $t_\perp$.  
With YBCO in mind, one of the two layers will be assumed to have 
tetragonal symmetry and model a CuO$_2$ plane, while the other will be 
taken to have orthorhombic symmetry and represent a CuO chain.
While this model is admittedly crude, it does allow us to examine the
role of interband transitions on the optical conductivity.  
We are interested in addressing two questions:  what is the magnitude of
the interband contribution to the conductivity compared to the
intraband (or Drude) contribtuion, and how different is the
frequency dependence of the interband contribution from that of the Drude
contribution?

Not surprisingly, we find that the interband contributions are of order
$t_\perp^2$ and are therefore relatively
unimportant for the $a$- and $b$-axis optical conductivities.  On the
other hand, the results are reversed for the $c$-axis conductivity:
the interband contributions are of order $t_\perp^2$ and dominate the
intraband contributions---which are of order $t_\perp^4$---for weak
interlayer couplings.  It is not surprising, then, that the $c$-axis
conductivity should have a non-Drude frequency dependence.  In the work
which follows, we will examine this frequency dependence and compare it
with experiment.

The paper is structured as follows.  In section \ref{lay}, 
general expressions for the optical conductivity of a system with two 
layers per unit cell are derived. 
In section \ref{chain} a specific model (the plane-chain model) which 
is suitable for YBCO is introduced.
Numerical results are given for the conductivity which we present separately 
for $a$- $b$- (along the chains) and $c$- (perpendicular to the planes) 
directions.  In section \ref{clean}, the expression for the conductivity
derived in the previous section is reduced analytically, with the intention
of highlighting the two types of contribution (interband and intraband) to
the conductivity.  One of the important results of this section is to
show how the different contributions to the conductivity depend on the
chain-plane coupling $t_\perp$.
In section \ref{bilayer}, expressions for the conductivity in the case of
a bilayer---consisting of two identical but unevenly space planes in
each unit cell---are derived.  The calculation is interesting because
YBCO, as well as many other cuprate superconductors, contains a CuO$_2$
bilayer in the unit cell.  One of our main conclusions in this section is
that the coupling between the CuO$_2$ planes is not as likely to be the
source of the broad background seen in the $c$-axis optical conductivity
as is the coupling between the CuO$_2$ planes and the chains.
Section \ref{sum} consists of a short discussion of sum rules.
A long conclusion, which includes some further discussion and a summary, is 
to be found in the final section.

\section{Conductivity in a layered system}
\label{lay}
The purpose of this section is to derive the equations needed for our
numerical calculations of the optical conductivity.  
In linear response theory, the real part of the conductivity tensor 
$\sigma_{\mu\nu}$
is related to the imaginary part of the current-current correlation 
function $\Pi_{\mu\nu}$ by \cite{Mahan}
\begin{equation}
\label{1}
\mbox{Re} [ \sigma_{\mu\nu}(\omega)] = 
\frac{2}{\omega} \mbox{Im} [ \Pi_{\mu\nu}(\omega) ]
\end{equation}
where the Greek subscripts refer to spatial components and $\omega$ is the
frequency.  The factor of 2 is to account for electron spins, which will
otherwise be ignored for the rest of this article.  In the superconducting
state, the spins are dealt with explicitly in the calculation of 
$\Pi(\omega)$.  In a previous
article we evaluated $\Pi$ for a two-layered tight binding system
\cite{Atkinson}.  The intention of our earlier calculations
was to find the penetration depth, so that $\Pi$ was evaluated 
at zero frequency ($\omega = 0$) and in the superconducting state.  Here
we will evaluate $\Pi(\omega)$ at finite frequency and in the normal
state.  As before, the model is a two-layer tight binding model, so
that the calculation is very similar to our earlier one.  For this
reason, the reader is referred to our earlier work for details of the
calculation which are not shown here.

In previous work, we showed that\cite{Atkinson}
\begin{eqnarray}
\label{2}
\Pi_{\mu\nu}(i\nu_n) &=& e^2 \frac{1}{\beta} \sum_m \frac{1}{\Omega} 
\sum_{\bf k}\mbox{Tr} \left [ G({\bf k};i\omega_m-i\nu_n) \right. \nonumber \\
&\times & \left. \gamma_\mu({\bf k},{\bf k})
G({\bf k};i\omega_m) \gamma_\nu({\bf k},{\bf k}) \right ]
\end{eqnarray}
where $\nu_n$ and $\omega_m$ are the boson and fermion Matsubara frequencies
respectively, $G({\bf k};i\omega_m)$ are the thermal Green's functions and
$\gamma_\mu$ are the electromagnetic
vertex functions.  This result is essentially the
standard result,\cite{Mahan} with the exception that here the Green's 
functions and vertex functions are $2 \times 2$ matrices whose diagonal
elements (eg.\ $G_{11}$) describe properties of a single layer and whose
off-diagonal elements describe the interlayer coupling.  The trace in
Eq.\ (\ref{2}) is over the matrix product contained in the square brackets.  
In order to find the optical conductivity, we need to find
explicit forms for $G$ and $\gamma_\mu$, and we begin by introducing 
our model for the two layer system.

The model we are going to present describes a metallic system with
two types of layer stacked along the $c$-axis 
(or equivalently, the $z$-axis).  In section \ref{chain} one of the
layers is a two-dimensional plane layer, while the other is a one-dimensional
chain layer.  In section \ref{bilayer} both layers are plane layers.
We define the operators $c_{1{\bf k}}$ and $c_{2{\bf k}}$ to be the 
annihilation operators for the two types of layer.  The wavevectors
${\bf k}$ are three-dimensional.  The Hamiltonian \cite{Atkinson,Tachiki,%
AYu} for our model is
\begin{equation}
\label{4}
H = \sum_{\bf k} \left[ \begin{array}{cc} 
c_{1{\bf k}}^\dagger & c_{2{\bf k}}^\dagger
\end{array} \right ] 
\, h({\bf k}) \,
\left[ \begin{array}{c} 
c_{1{\bf k}} \\ c_{2{\bf k}}
\end{array} \right ],
\end{equation}
with
\begin{equation}
h({\bf k}) = 
\left[ \begin{array}{cc} 
\xi_1 & t \\ t^\ast & \xi_2
\end{array} \right ],
\end{equation}
and where $\xi_1({\bf k})$ and $\xi_2({\bf k})$ are the energy dispersions
for the two types of layer, and $t({\bf k})$ connects the layers through
single electron hopping.  The specific forms of $\xi_1$, $\xi_2$ and $t$
will be given in sections \ref{chain} and \ref{bilayer}.

The Hamiltonian matrix $h({\bf k})$ is diagonalised by the unitary matrices 
$U({\bf k})$, so that
\begin{equation}
\left [ \begin{array}{cc}
\epsilon_+({\bf k}) & 0 \\ 0 & \epsilon_-({\bf k}) \end{array} \right ] = 
U^\dagger({\bf k}) h({\bf k}) U({\bf k}),
\end{equation}
where
\begin{equation}
U({\bf k}) = \frac{1}{\sqrt{\epsilon_+ - \epsilon_-}}
\left [ \begin{array}{cc}
-\frac{t}{|t|}\sqrt{\xi_1 - \epsilon_-} & -\frac{t}{|t|}
\sqrt{\epsilon_+ - \xi_1} \\[.5cm] 
-\sqrt{\epsilon_+ - \xi_1} & \sqrt{\xi_1 - \epsilon_-} 
\end{array} \right ],
\end{equation}
and where $\epsilon_\pm({\bf k})$ are the eigenvalues of $h({\bf k})$ (ie.\
$\epsilon_\pm({\bf k})$ are the band energies),
\begin{equation}
\label{6}
\epsilon_\pm = \frac{\xi_1+\xi_2}{2} \pm \sqrt{\left(\frac{\xi_1-\xi_2}{2}
\right )^2 + t^2}.
\end{equation}
The Fermi surfaces are the solutions of the equations $\epsilon_\pm({\bf k})
= 0$.  As an example, one possible Fermi surface for the chain-plane system%
---which is discussed in more detail in section \ref{chain}---is
shown in Fig.\ \ref{f1}.  


The single particle Green's function is determined from the Hamiltonian:
\begin{equation}
\label{7}
G({\bf k};i\omega_n)^{-1} = \left [ \begin{array}{cc}
i\omega_n -\xi_1 & -t \\ -t^\ast & i\omega_n - \xi_2 \end{array} \right ].
\end{equation}
If we wish to include impurity scattering, then the simplest approach
is to introduce a scattering rate $\Gamma$ 
\begin{eqnarray}
\label{8}
&&G({\bf k};i\omega_n)^{-1} = \nonumber \\
&&\left [ \begin{array}{cc}
i\omega_n -\xi_1 +i\Gamma \mbox{sgn}(\omega_n)  & -t \\ 
-t^\ast & i\omega_n - \xi_2 +i\Gamma \mbox{sgn}(\omega_n) 
\end{array} \right ].
\end{eqnarray}
We will assume that $\Gamma$ is independent of frequency and momentum,
but that it varies linearly with temperature as is observed in 
the copper oxides for the in-plane conductivity.  At $T=0$, the system
is in the clean limit ($\Gamma = 0$) and at $T=100$ K, $\Gamma = 10$ meV.
The scattering rate is related to the quasiparticle lifetime by
$\Gamma = \hbar/2\tau$.

Finally, we need to find the vertex function $\gamma_\mu({\bf k},{\bf k})$.  
In previous work\cite{Atkinson,Hirsch} it has been shown that 
in the tight binding
model, $\gamma_\mu({\bf k},{\bf k})$ is just the gradient of the Hamiltonian
matrix in ${\bf k}$ space:
\begin{equation}
\label{10}
\gamma_\mu({\bf k},{\bf k}) = \frac{1}{\hbar} \frac{\partial}{\partial k_\mu}
\left[ \begin{array}{cc} 
\xi_1 & t \\ t^\ast & \xi_2
\end{array} \right ].
\end{equation}

Equations (\ref{2}), (\ref{8}) and (\ref{10}) are sufficient to calculate
the optical conductivity.  In their current form, however, they are not
very revealing and it is difficult to understand the results of our 
numerical calculations without some further work.  We will make two
manipulations in order to make the formula for the conductivity more
transparent.  The first is to write the conductivity in terms of
spectral functions, instead of Green's functions:
\begin{eqnarray}
\label{3}
\mbox{Re}[\sigma_{\mu\nu}(\omega)] &=& \frac{e^2 \hbar}{2\pi \Omega} 
\sum_{\bf k} \int_{-\infty}^\infty dx \,
\mbox{Tr} \left [ A({\bf k};x) \gamma_\mu({\bf k},{\bf k}) \right. \nonumber \\
&\times& \left. A({\bf k};x+\hbar \omega) \gamma_\nu({\bf k},{\bf k}) \right ] 
\frac{ f(x) - f(x+\hbar\omega)}{\hbar \omega}, \nonumber \\
\end{eqnarray}
where the spectral function $A({\bf k};\omega)$ is defined by analytically 
continuing $G({\bf k},i\omega_n)$ to the real axis:
\begin{displaymath}
A({\bf k};\omega) = i \left[ G({\bf k};\omega+i0) -
 G({\bf k};\omega-i0) \right]
\end{displaymath}
and where $f(x)$ is the Fermi function. 

The second manipulation is to make a change of basis,
so that the eigenstates of the Hamiltonian, rather than 
the eigenstates of the isolated layers, are used as the basis states.
In other words, we will evaluate
\begin{eqnarray}
\label{6a}
\mbox{Re}[\sigma_{\mu\nu}(\omega)] &=& \frac{e^2 \hbar}{2\pi\Omega} 
\sum_{\bf k} \int_{-\infty}^\infty dx \,
\mbox{Tr} \left [ \hat{A}({\bf k};x) \hat{\gamma}_\mu({\bf k},{\bf k}) 
\right. \nonumber \\
&\times& \left. \hat{A}({\bf k};x+\hbar \omega) \hat{\gamma}_\nu({\bf k},
{\bf k}) \right ] \frac{ f(x) - f(x+\hbar\omega)}{\hbar \omega}, \nonumber \\
\end{eqnarray}
where we introduce the notation throughout this work
$\hat{O}({\bf k}) \equiv U^\dagger({\bf k}) O({\bf k}) U({\bf k})$,
with $O({\bf k})$ a $2\times 2$ matrix.

Equation (\ref{6a}) is more useful than Eq.\ (\ref{2}) for two reasons.
The first is that the spectral function has a simple interpretation
as the density of states.  The second is that $\hat{A}$ has a particularly
simple form.  The price we pay is that the vertex function $\hat{\gamma}_\mu$ 
is more complicated to evaluate than $\gamma_\mu$.

The vertex function $\hat{\gamma}_\mu$ can be found by 
explicitly performing the matrix multiplication $U^\dagger({\bf k})
\gamma_\mu U({\bf k})$, and we will explore it in detail in the
following sections.  The spectral function can be evaluated easily
here.  In the new basis
\begin{eqnarray}
&&\hat{G}({\bf k};i\omega_n)^{-1} = \nonumber \\
&&\left [ \begin{array}{cc}
i\omega_n -\epsilon_+ +i\Gamma \mbox{sgn}(\omega_n)  & 0 \\ 
0 & i\omega_n - \epsilon_- + i\Gamma \mbox{sgn}(\omega_n) 
\end{array} \right ],
\end{eqnarray}
and 
\begin{equation}
\label{9}
\hat{A}({\bf k};\omega) = \left [ \begin{array}{cc}
\frac{2\Gamma}{(\omega-\epsilon_+)^2 + \Gamma^2} 
& 0 \\[.5cm]
0 & \frac{2\Gamma}{(\omega-\epsilon_-)^2 + \Gamma^2} .
\end{array} \right ]
\end{equation}
The diagonal elements of the spectral function can be interpreted as
the density of electronic states in the bands.

\section{Plane-chain model}
\label{chain}

In this section we describe a simple model containing a plane layer and
a chain layer.  The model is meant to incorporate the most important features
of the chain-plane coupling in YBCO$_{7-\delta}$.  In the discussion which
follows, it will be made clear that the interband terms 
in the optical conductivity (which are important for $\sigma_{zz}$) are 
sensitive to the specifics of the band structure, which we cannot hope 
to describe correctly with our model.  
However, we will still able to draw a number
of general conclusions which should apply to models with a more accurate
description of the unit cell.

For this work we will take the dispersion in the plane layer to be
\begin{mathletters}
\label{5}
\begin{eqnarray}
\label{5a}
\xi_1 &=& -2t_1 [\cos(k_xa) + \cos(k_ya) \nonumber \\
&& -2B\cos(k_xa)\cos(k_ya) ] - \mu_1,
\end{eqnarray}
and the dispersion in the chain layer to be
\begin{equation}
\label{5b}
\xi_2 = -2t_2 \cos(k_ya) - \mu_2,
\end{equation}
where $a$ is the lattice constant in the $a$ (or equivalently $x$) and
$b$ (or $y$) directions.
There have been a number of attempts to fit the Fermi surface of the
plane layer either
to band structure calculations \cite{Andersen,Pickett,JYu} or to angle 
resolved 
photoemission experiments \cite{Blumberg}, however, we are unaware of
any work that attempts to find a phenomenological chain Fermi surface.
In any case, a more accurate Fermi surface will not affect our basic
(qualitative) conclusions, although the quantitative results of our 
calculations are quite sensitive to the band structure.

The simplest model of interlayer coupling is
\begin{equation}
\label{5c}
t = -2t_\perp \cos(k_zd/2),
\end{equation}
\end{mathletters}
where $d$ is the lattice constant in the $c$ (or $z$) direction.
The form of Eq.\ (\ref{5c}) makes the assumption that we have 
{\em coherent} single electron transport along the $c$-axis.  

There have been many models of incoherent transport between 
layers.\cite{Zha,Graf,Kumar,Rojo,Leggett,Anderson} A review of some of the 
models has been provided by Cooper and Gray.\cite{Cooper}  
As these models are not directly relevant to our present work, it will be 
sufficient here to provide a few highlights.  
In the work of Graf et al.,\cite{Graf} no contribution from coherent 
transport in envisaged (i.e. $t_\perp=0$) and the $c$-axis transport 
proceeds entirely through incoherent elastic scattering.  In this case, the 
resistivity along $c$- is inversely proportional to that in the $ab$-plane.  
On the other hand, Kumar and Jayannavar\cite{Kumar} envisage tunnelling 
between layers but in the limit that the tunnelling time is larger than the 
in-plane scattering time. 
The effective transverse tunnelling matrix element is modulated and 
reduced by the in-plane scattering and the $c$-axis conductivity becomes 
proportional to the $a-b$ plane scattering time.  Leggett\cite{Leggett} 
envisages thermal depairing between 
layers for the case $t_\perp<k_B T$ while Anderson and Zhou\cite{Anderson} 
consider the possibility that the CuO$_2$ planes cannot be described in Fermi
liquid theory and involve spin and charge separation.  This implies 
blocking of $c$-axis transport.  Finally, 
Rojo and Levin\cite{Rojo} consider the possibility that $c$-axis 
transport proceeds through the combination of $t_\perp$ and
incoherent transport due to elastic (impurity)
and inelastic (phonon) assisted processes.

Here only the coherent contribution is included.  It should also be remembered
that we have oversimplified the internal structure of the unit cell in YBCO,
particularly since we have only included one type of hopping mechanism
which is paramaterised by $t_\perp$.  In reality, there could be several
types of hopping process occuring.  For example, the coupling 
between planes and chains could be coherent while the intercell coupling 
could be incoherent.  These issues are not treated here and go beyond the 
scope of our work.  They are mentioned, however, so that the reader 
understands clearly the limitations of our work.  We do feel, however, that
near optimal doping YBCO is one of the few materials to display coherent
transport between layers, and that our description of the interlayer
coupling is reasonable. 

The vertex functions are
\begin{mathletters}
\begin{equation}
\gamma_x({\bf k},{\bf k}) = 
\left[ \begin{array}{cc} 
v_{1x} &0 \\ 0 & 0 
\end{array} \right ],
\end{equation}
\begin{equation}
\gamma_y({\bf k},{\bf k}) =
\left[ \begin{array}{cc} 
v_{1y} & 0 \\ 0 & v_{2y}
\end{array} \right ],
\end{equation}
and
\begin{equation}
\gamma_z({\bf k},{\bf k}) = 
\left[ \begin{array}{cc} 
0 & v_\perp \\ v_\perp & 0 
\end{array} \right ].
\end{equation}
\end{mathletters}
where $v_{i\mu} = \hbar^{-1} {\partial \xi_i}/{\partial k_\mu}$,
and $v_\perp = \hbar^{-1} {\partial t}/{\partial k_z}$.

The change of basis is simple to perform and
\begin{mathletters}
\label{11}
\begin{equation}
\label{11a}
\hat{\gamma}_x({\bf k},{\bf k}) = 
\frac{v_{1x}}{\epsilon_+ - \epsilon_-}
\left[ \begin{array}{cc} 
\xi_1 - \epsilon_- & t \\ t & \epsilon_+ - \xi_1
\end{array} \right ],
\end{equation}
\begin{eqnarray}
\label{11b}
\hat{\gamma}_y({\bf k},{\bf k}) &=& 
\frac{v_{1y}}{\epsilon_+ - \epsilon_-}
\left[ \begin{array}{cc} 
\xi_1 - \epsilon_- & t \\ t & \epsilon_+ - \xi_1
\end{array} \right ] \nonumber \\[.4cm]
&+& \frac{v_{2y}}{\epsilon_+ - \epsilon_-} \left[ \begin{array}{cc} 
\xi_2 - \epsilon_- & t \\ t & \epsilon_+ - \xi_2
\end{array} \right ],
\end{eqnarray}
and
\begin{equation}
\label{11c}
\hat{\gamma}_z({\bf k},{\bf k}) = \frac{v_\perp}{\epsilon_+ - \epsilon_-}
\left[ \begin{array}{cc} 
2t & \xi_2 - \xi_1 \\ \xi_2 - \xi_1 & 2t
\end{array} \right ].
\end{equation}
\end{mathletters}
In the derivation of Eqs.\ (\ref{11}), use was made of the fact that
$t({\bf k}) \geq 0$.
We are now able to evaluate the real part of the conductivity tensor
[Eq.\  (\ref{6a})] using Eqs.\ (\ref{9}) and (\ref{11}).

In Fig.~\ref{f2} the conductivity is plotted as a function of frequency
for a number of different temperatures.  The strength of the interlayer
coupling is relatively weak ($t_\perp = 5$ meV) and this is reflected
in the small magnitude of the $c$-axis conductivity.  
We point out that the case presented in Fig.~\ref{f2} provides 
a qualitative description of what is observed underdoped 
YBCO$_{6.7}$.\cite{Homes,Homes2}  To begin with, the 
conductivity in the $a$ direction has a traditional Drude-like structure.
Since the scattering rate scales linearly with temperature, the width
of the Drude-peak also scales linearly with temperature, and at $T=0$,
the conductivity is a $\delta$-function at $\omega = 0$.  The conductivity
in the $b$ direction also has a predominantly Drude-like structure but its
magnitude is roughly twice that of $\sigma_{xx}$.  This is because there
are two current carrying channels in the $b$ direction (the chains and the
planes) and only one in the $a$ direction (the planes).  At 
low temperatures there is a small non-Drude contribution
due to interband transitions which becomes visible, and at $T=0$, the
interband transitions are the only mechanism for conductivity at 
finite frequency.

In contrast with $\sigma_{xx}(\omega)$ and $\sigma_{yy}(\omega)$,
the conductivity
in the $c$ direction has a decidedly non-Drude behaviour and, instead,
appears as a broad background.  As we
shall see, this is entirely due to interband transitions between the
plane and chain layers.  A further, interesting feature in $\sigma_{zz}
(\omega)$
is that at high temperatures there is very little structure as a function
of frequency $\omega$, but that structure appears at low temperatures.
This can be attributed to the linear decrease in the scattering rate
with temperature assumed in our work; at high temperature the large scattering rate smears
out the structure in $\sigma_{zz}(\omega)$.
Perhaps the most striking feature in $\sigma_{zz}$ is the development
of a gap as the temperature is lowered.  This gap is just the gap between
the two bands $\epsilon_+$ and $\epsilon_-$, and at high temperatures it
is filled in by the large scattering rate $\Gamma$.

In Figs.~\ref{f3} and \ref{f4}, the optical conductivity is plotted
for larger values of the chain-plane coupling $t_\perp$.  The larger
coupling has almost no effect on $\sigma_{xx}$.  On the other hand,
the larger coupling enhances the role of interband transitions in
$\sigma_{yy}$, and also increases the Drude (or intraband) contribution
to $\sigma_{zz}$.  Note that the Drude peak at low $\omega$ is not seen in 
$\sigma_{zz}$ in Fig. ~\ref{f2} where $t_\perp$ is small, 
while in Fig.~\ref{f4}, where $t_\perp$ is much larger, the Drude peak 
is seen as a sharp turn upward at low $\omega$.  

The conductivities shown in Figs.~\ref{f3} and \ref{f4}
are not quite appropriate to describe optimally doped YBCO even though the
magnitudes of the conductivities are approximately right.  This is because
there is still a gap in the conductivity at low temperatures.  However,
by shifting the chain Fermi surface slightly, as shown in Fig.~\ref{f5},
we are able to eliminate the band gap.  It is very plausible that a 
change in doping will change the structure of the chain and plane Fermi 
surfaces so that they cross.  In Figs.~\ref{f6} and \ref{f7} the conductivity
is shown at $T = 100$~K and at $T=0$~K for $t_\perp$=10 meV and 20 meV, respectively.  In Fig.~\ref{f6}, the 
conductivity has a Drude-like appearance in both the $a$ and $b$ directions,
even though there is a large interband contribution to $\sigma_{yy}$.
The interband contribution can be resolved in $\sigma_{yy}$ if the 
temperature (and therefore the scattering rate) is reduced.  Unfortunately,
this is difficult to do experimentally because of the onset of 
superconductivity near 100 K.  In both Fig.~\ref{f6} and \ref{f7}, 
the $c$-axis response shows a Drude-like peak at low $\omega$ and 
high temperature (solid curve).  On closer examination, the response is a 
combination of a Drude and flat interband contribution.  As the temperature 
is lowered, this second contribution remains (dotted curve) and a 
pseudogap at low $\omega$ is resolved 
(i.e. the conductivity is depressed but goes strictly to zero only at 
$\omega$=0).

\section{Clean Limit}
\label{clean}
The purpose of this section is to understand the role of interband
transitions a little better.  We will examine why it is that the
interband transitions are most important in the $c$-axis conductivity
for small interlayer coupling, while they are most important for 
the $b$-axis conductivity for large interlayer coupling, and are never 
important for the $a$-axis conductivity.
Finally, we shall derive expressions for the intraband and interband
conductivities in the clean ($\Gamma = 0$) limit and show how
$\sigma_{zz}(\omega)$ can be interpreted as a probe of the band structure.

Let us start by writing out
Eq.\ (\ref{6a}), the formula for the conductivity, explicitly:
\widetext
\begin{eqnarray}
\label{16}
\mbox{Re}[\sigma_{\mu\mu}(\omega)] &=& 
\frac{e^2 \hbar}{2\pi\Omega} 
\sum_{\bf k} \int_{-\infty}^\infty dx \,
\left [ \hat{A}_{11}({\bf k};x) \hat{A}_{11}({\bf k};x+\hbar \omega) 
[\hat{\gamma}_{11}]_\mu^2 
+ \hat{A}_{22}({\bf k};x) \hat{A}_{22}({\bf k};x+\hbar \omega) 
[\hat{\gamma}_{22}]_\mu^2 \right . \nonumber \\
&+& \left.\left 
( \hat{A}_{11}({\bf k};x) \hat{A}_{22}({\bf k};x+\hbar \omega) 
+ \hat{A}_{22}({\bf k};x) \hat{A}_{11}({\bf k};x+\hbar \omega) \right )
[\hat{\gamma}_{12}]_\mu [\hat{\gamma}_{21}]_\mu \right ]
\frac{ f(x) - f(x+\hbar\omega)}{\hbar \omega},
\end{eqnarray}
\narrowtext
The first two terms in the integrand describe {\em intraband} processes
and yield Drude-like behaviour.  This fact can be made clear if we
remark that
\begin{equation}
\label{10a}
[\hat{\gamma}_{ij}]_\mu = \frac{\delta_{ij}}{\hbar} \frac{\partial 
\epsilon_i}{\partial k_\mu} + \frac{1}{\hbar} [\epsilon_i - \epsilon_j]
\left[ U^\dagger({\bf k}) \frac{\partial U({\bf k})}{\partial k_\mu} 
\right ]_{ij},
\end{equation}
(where we use the notation $\epsilon_1 \equiv \epsilon_+$ and
$\epsilon_2 \equiv \epsilon_-$)
so that the first two terms in the conductivity are
\begin{eqnarray}
\sigma_{\mbox{\scriptsize Drude}}(\omega) &=&
\frac{e^2 \hbar}{2\pi\Omega} \sum_\pm \sum_{\bf k} \int_{-\infty}^\infty dx \,
\left[ \frac{1}{\hbar} \frac{\partial \epsilon_\pm}{\partial k_\mu} \right ]^2
\nonumber \\
&&\times \frac{2\Gamma}{(x-\epsilon_\pm)^2 + \Gamma^2} 
\frac{2\Gamma}{(\hbar\omega + x-\epsilon_\pm)^2 + \Gamma^2}
\nonumber \\
&&\times \frac{f(x)-f(x+\hbar\omega)}{\hbar\omega} . 
\end{eqnarray}
In the simple case of a single band system with a spherical Fermi surface,
this becomes the usual expression for the optical conductivity
\[
  \sigma(\omega) = 2 e^2 N(0) \frac{v_F^2}{3} \frac{\tau}{1+\omega^2\tau^2},
\]
where $\Gamma = \hbar/2\tau$.

In the clean limit ($\Gamma \rightarrow 0$) 
\[
\frac{\Gamma}{x^2 + \Gamma^2} \rightarrow \pi \delta(x).
\]
and the intraband conductivity becomes
\begin{equation}
\label{17a}
\sigma_{\mbox{\scriptsize Drude}}(\omega) = 
- \frac{2\pi e^2 \hbar}{\Omega} \sum_\pm \sum_{\bf k} 
\left[ \frac{1}{\hbar} \frac{\partial \epsilon_\pm}{\partial k_\mu} \right ]^2
\frac{\partial f(\epsilon_\pm)}{\partial \epsilon_\pm} \delta(\hbar\omega).
\end{equation}
The Drude peak in the conductivity is now a $\delta$-function centred at
$\omega=0$.

The remaining terms in Eq.\ (\ref{16}) are the {\em interband} terms. 
In the clean limit the interband contribution is (for $\omega > 0$)
\begin{eqnarray}
\label{18}
\sigma_{\mbox{\scriptsize Inter}}(\omega) & = &
\frac{2 \pi e^2 \hbar}{\Omega} 
\sum_{\bf k}  \left [ \hat{\gamma}_{12} \right ]_\mu^2
\frac{f(\epsilon_-)-f(\epsilon_+)}{\epsilon_+-\epsilon_-} \nonumber \\
&&\times \delta(\hbar \omega - \epsilon_+ + \epsilon_-). 
\end{eqnarray}

In order to get some sense of the relative magnitudes of 
$\sigma_{\mbox{\scriptsize Drude}}(\omega)$ and
$\sigma_{\mbox{\scriptsize Inter}}(\omega)$ we compare the magnitudes of
$[ \hbar^{-1} {\partial \epsilon_\pm}/{\partial k_\mu} ]^2$ and
$[ \gamma_{12} ]_\mu^2$.  In the analysis which follows, we will
compare these two quantities for $\sigma_{xx}$,
$\sigma_{yy}$ and $\sigma_{zz}$.

In the $x$-direction,
\[
\left [ \frac{1}{\hbar} \frac{\partial \epsilon_\pm}{\partial k_x} 
\right ]^2 =
\left [ v_{1x} \frac{\xi_1 - \epsilon_\mp}{\epsilon_+ - \epsilon_-} 
\right ]^2 
\]
and
\[
[ \hat{\gamma}_{12} ]_x^2 =
\left [ v_{1x} \frac{t}{\epsilon_+ - \epsilon_-}\right ]^2 .
\]
Throughout most of the Brillouin zone, $|\xi_1 - \xi_2| \gg 2|t|$ and
\[
\epsilon_+ \sim \max(\xi_1,\xi_2) + \frac{t^2}{(\xi_1-\xi_2)^2}
\]
\[
\epsilon_- \sim \min(\xi_1,\xi_2) - \frac{t^2}{(\xi_1-\xi_2)^2}
\]
from which it follows that the largest contribution to 
$\sigma_{\mbox{\scriptsize Drude}}(\omega)$ will be of the order
$v_{1x}^2 \times [1 + O(t^2/(\xi_1-\xi_2)^2)]$, 
where the correction term is of
the order 1/25 throughout most of the Brillouin zone.  In the same way,
we can immediately see that throughout most of the Brillouin zone,
the interband contribution to $\sigma_{xx}$ will be $v_{1x}^2 \times
O(t^2/(\xi_1-\xi_2)^2)$.  Of course, in the region of the Brillouin zone
where the chain and plane Fermi surfaces are close together, the above 
argument does not hold.  However, the value of $v_{1x}$ is small in
this region of the Brillouin zone and we conclude that interband processes 
do not make a significant contribution to $\sigma_{xx}$.  

Much of the above argument holds for $\sigma_{yy}$ as well, and
we can conclude that throughout most of the Brillouin zone, interband
processes are unimportant.  Unlike the case of $\sigma_{xx}$, however,
$v_{1y}$ and $v_{2y}$ are not small in regions of the Brillouin zone
where interband transitions are significant, and there will therefore
be small but noticeable non-Drude contribution to the $b$-axis
conductivity.

The situation is somewhat reversed for the $c$ axis conductivity, where
we can show that interband processes play a dominant role.
In the $z$-direction
\[
\left [ \frac{1}{\hbar} \frac{\partial \epsilon_\pm}{\partial k_z} 
\right ]^2 =
\left [ v_{\perp} \frac{2t}{\epsilon_+ - \epsilon_-} \right ]^2 
\]
and
\[
[ \hat{\gamma}_{12} ]_z^2 =
\left [ v_{\perp} \frac{\xi_1-\xi_2}{\epsilon_+ - \epsilon_-}\right ]^2 .
\]
Inspection of these two equations shows that 
while the interband contribution still scales as
$t_\perp^2$ (through the factor of $v_\perp$), the Drude component
now scales as $t_\perp^4$.  In other words, Drude contribution to the 
conductivity is
smaller than the interband contribution by a factor $t^2/(\xi_1-\xi_2)^2$.

In summary, then, the conductivities scale with
$t_\perp$ as follows:  For $\sigma_{xx}$ and $\sigma_{yy}$, we find
that the Drude part scales as $(t_\perp)^0$ and the interband part scales
as $t_\perp^2$, while for $\sigma_{zz}$ the interband part scales as
$t_\perp^2$ and the Drude part as $t_\perp^4$.

We will finish this section with a brief mention of the usefulness
of $\sigma_{zz}$ as a probe of the band structure.
The factor $\delta(\hbar \omega - \epsilon_+ + \epsilon_-)$ 
in Eq.\ (\ref{18}) means that the interband conductivity
is a probe of the {\em joint density of states} of the two bands.
The sum in Eq.\ (\ref{18}) is weighted by the thermal factor
$f(\epsilon_-) - f(\epsilon_+)$, which restricts the
transitions to be between filled and empty states.  
At zero temperature, this term restricts
the sum to lie in regions of the Brillouin zone where $\epsilon_+
\epsilon_- < 0$ (ie.\ $\epsilon_-({\bf k})$ is a filled state and
$\epsilon_+({\bf k})$ is an empty state).  
In Fig.\ \ref{f1}, this means that the interband
transitions occur in the area contained by the two Fermi surface
curves.  The constant energy-difference contours are also shown
in Fig.~\ref{f1} as the dotted lines.

When we compare the chain-plane system with the bilayer model in the
next section, we will see that the chain-plane system is somewhat special
in that the joint density of states is spread over a broad range of energies.
In simple terms, the interband contribution to the $c$-axis conductivity
exists over an energy scale of 1 eV because the chain and plane bands
have such different structure, and the energy difference $\epsilon_+({\bf k})
-\epsilon_-({\bf k})$ takes on all different values on the energy scale
of an eV in
the Brillouin zone.  In the bilayer system, however, where the two layers
are identical, the energy difference is restricted to a narrow range of 
values, and the interband contribution to the optical conductivity will
result in a narrow peak.

\section{Bilayer model}
\label{bilayer}

In this model, we have two planes per unit cell.  The planes have equivalent
band structures, but are made inequivalent by their spacing.  In other words,
the planes are alternately spaced by distances $d_1$ and $d_2$ along
the $c$-axis, where $d_1 + d_2 = d$ is the unit cell length.  In the
special case $d_1 = d_2$, the model reduces to a single band model as
we shall show below.  

The energy dispersion in the planes is $\xi_1 = \xi_2 = \xi$.  For our
numerical calculations, we will take
\begin{mathletters}
\begin{eqnarray}
\label{13a}
\xi &=& -2t_1 [\cos(k_xa) + \cos(k_ya) \nonumber \\
&& -2B\cos(k_xa)\cos(k_ya) ] - \mu_1.
\end{eqnarray}
In this section, we will also calculate the conductivity analytically
using the simpler band structure
\begin{equation}
\label{13c}
\xi = \frac{\hbar^2}{2m} (k_\|^2 - k_F^2)
\end{equation}
where $v_F$ is the Fermi velocity at the (circular) Fermi surface, 
$k_\| = \sqrt{k_x^2+k_y^2}$ and
$k_F$ is the value of $k_\|$ at the Fermi surface.

The interlayer coupling term takes the form\cite{Atkinson}
\begin{equation}
\label{13b}
t({\bf k}) = t_{\perp 1}e^{ik_zd_1} + t_{\perp 2} e^{-ik_zd_2},
\end{equation}
\end{mathletters}
where we expect that if $d_1 < d_2$, then $t_{\perp 1} > t_{\perp 2}$.
If $d_1 = d_2$ and $t_1 = t_2$, then $t({\bf k})$ reduces to Eq.\ (\ref{5c}).
The band energies are
\begin{equation}
\epsilon_\pm = \xi({\bf k}) \pm |t({\bf k})|,
\end{equation}
and the Fermi surface is shown in Fig.\ \ref{f9} for the case where
$\xi$ is given by Eq.\ (\ref{13a}).  In the bi-planar model, the
bands are split by $2|t({\bf k})|$ so that the maximum band energy difference
is $2|t_{\perp 1} + t_{\perp 2}|$ and the minimum band energy difference
is $2|t_{\perp 1} - t_{\perp 2}|$.  Since $t_{\perp 1}$ and $t_{\perp 2}$
are typically much smaller than the bandwidths, this means that the
interband contribution to $\sigma_{zz}$ will be over a small range of
energies.  As we will see from our numerical work, the bilayer is not
a likely source for the broad, experimentally observed, $c$-axis response.

The unitary matrix which diagonalises the Hamiltonian is now
\begin{equation}
U({\bf k}) = \frac{1}{\sqrt{2}|t|} \left [ \begin{array}{cc}
-t & -t \\ -|t| & |t| \end{array} \right].
\end{equation}

The vertex function for the in-plane conductivity  $\sigma_{xx}$ is
\begin{equation}
\label{14a}
\hat{\gamma}_x({\bf k},{\bf k}) = \gamma_x({\bf k},{\bf k}) = 
\left[ \begin{array}{cc} 
v_x & 0 \\ 0 & v_x
\end{array} \right ],
\end{equation}
where $ v_x = \hbar^{-1} {\partial \xi}/{\partial k_x}$.
We can see that there will be no interband contribution to
$\sigma_{xx}(\omega)$ since $\hat{\gamma}_x$ has no off-diagonal
matrix elements.  In fact, Eq.\ (\ref{6a}) becomes 
\begin{eqnarray}
\mbox{Re}[\sigma_{xx}(\omega)] &=&
\frac{e^2\hbar}{2\pi\Omega} \sum_\pm \sum_{\bf k} 
\int_{-\infty}^{\infty} dx \, 
 v_x^2 \, \frac{f(x)-f(x+\hbar\omega)}{\hbar\omega}  \nonumber \\
&&\times \frac{2\Gamma}{(x-\epsilon_\pm)^2 + \Gamma^2}
\frac{2\Gamma}{(x+ \hbar \omega-\epsilon_\pm)^2 + \Gamma^2}. \nonumber \\
\end{eqnarray}
If we take $\xi$ from Eq.\ (\ref{13c}) then we get, in the
usual way, the Drude conductivity
\begin{equation}
\label{20}
\mbox{Re}[\sigma_{xx}] = \frac{4 e^2 \hbar N_\|}{d} 
\frac{v_F^2}{2} \frac{\tau}{\omega^2\tau^2+1},
\end{equation}
where $N_\|$ is the two dimensional density of states for a single layer.
In three dimensions, with a cylindrical Fermi surface, the density of
states for the bilayer is $N = 2N_\|/d$.  
Numerical calculations of the optical conductivity, shown in Fig.~\ref{f10},
are in qualitative agreement with Eq.~(\ref{20}).

The vertex function for the conductivity along the $c$-axis is
\begin{eqnarray}
\label{14b}
\hat{\gamma}_z({\bf k},{\bf k}) &=& 
U^\dagger({\bf k}) \left[ \begin{array}{cc} 
0 & v_\perp \\ 
v_\perp^\ast & 0
\end{array} \right ] U({\bf k}) \nonumber \\
&=& \frac{1}{|t|} \left [ \begin{array}{cc}
\mbox{Re} (v_z t^\ast) & -i \mbox{ Im}(v_z t^\ast) \\
i \mbox{ Im}(v_z t^\ast) & - \mbox{ Re} (v_z t^\ast)
\end{array} \right ],
\end{eqnarray}
where $v_z = \hbar^{-1} {\partial t}/{\partial k_z}$.
The diagonal elements in $\hat{\gamma}_z$ are
\begin{eqnarray}
  \frac{\mbox{Re}(v_z t^\ast)}{|t|} &=& \frac{1}{\hbar}
\frac{\partial |t|}{\partial k_z} \nonumber \\
   &=& \frac{1}{\hbar} 
  \frac{t_{\perp 1}t_{\perp 2}}{|t|} \, d \, \sin(k_z d) \nonumber \\
  &\leq& \frac{1}{\hbar}
  \min(t_{\perp 1},t_{\perp 2})  \, d \, \sin(k_z d), \nonumber 
\end{eqnarray}
so that the intraband (or Drude) conductivity is limited by the lessor of
$t_{\perp 1}$ and $t_{\perp 2}$.  In other words
the Drude current is limited by the weak link along the $c$-axis.
The off-diagonal elements in $\hat{\gamma}_z$ are
\begin{eqnarray}
\frac{\mbox{Im}({v_z t^\ast})}{|t|} &=&  \frac{1}{\hbar}
[t_{\perp 1}^2 d_1 - t_{\perp 2}^2 d_2 \nonumber \\
&&+ t_{\perp 1} t_{\perp 2}(d_1 -d_2) \cos(k_z d)]/|t|,
\end{eqnarray}
which vanishes when $t_1 = t_2$ and $d_1 = d_2$.  Unlike the intraband
conductivity, the interband conductivity does not become small as either
$t_{\perp 1}$ or $t_{\perp 2}$ vanishes.  In other words, the interband
contribution to $\sigma_{zz}$ persists even in the limit that the
bilayers become isolated from their neighbours.  

The conductivity along the $c$-axis is
\widetext
\begin{eqnarray}
\mbox{Re} [\sigma_{zz}(\omega) ] &=& \frac{e^2\hbar}{2\pi\Omega} 
\sum_{\bf k} \int_{-\infty}^{\infty} dx 
\, \frac{f(x)-f(x+\hbar\omega)}{\hbar\omega} 
\left \{\sum_\pm \frac{2\Gamma}{(x-\epsilon_\pm)^2 + \Gamma^2}
\frac{2\Gamma}{(x+ \hbar\omega-\epsilon_\pm)^2 + \Gamma^2} 
\left [ \frac{\mbox{Re}[v_zt^\ast]}{|t|} \right ]^2 \right . \nonumber \\
&& + \left. \sum_\pm \frac{2\Gamma}{(x-\epsilon_\pm)^2 + \Gamma^2}
\frac{2\Gamma}{(x+ \hbar\omega-\epsilon_\mp)^2 + \Gamma^2} 
\left [ \frac{\mbox{Im}[v_zt^\ast]}{|t|} \right ]^2 \right\}. 
\end{eqnarray}
The first term in the curly braces gives the intraband conductivity
while the second gives the interband conductivity.  We can proceed
further if we take $\xi$ to be of the form given in Eq.\ (\ref{13c}).
Then we find that
\begin{eqnarray}
\label{46}
\mbox{Re} [\sigma_{zz}(\omega) ] &=& \frac{2 e^2 N_\|}{d}
 \left \{ 2 \, \frac{\tau}{\omega^2\tau^2 + 1}
\left \langle
\left [ \frac{\mbox{Re}[v_zt^\ast]}{|t|} \right ]^2 \right \rangle_{k_z}
+ \ \sum_\pm \left \langle  \frac{\tau}
{\tau^2(\omega \pm 2|t|/\hbar)^2 + 1}
\left [ \frac{\mbox{Im}[v_zt^\ast]}{|t|} \right ]^2 \right \rangle_{k_z}
\right \},
\end{eqnarray}
\narrowtext
\noindent
where $\langle \rangle_{k_z}$ denotes an average over $k_z$.  

The first term in Eq.~(\ref{46}) has the usual Drude frequency dependence,
weighted by an average Fermi velocity.  The second term is the interband
term.  Its frequency dependence is an average over $k_z$ of Lorentzians 
centred at $\hbar \omega = 2|t(k_z)|$.  This is what is seen in 
Fig.~(\ref{f10}).  In the limit that we have a single
bilayer ($t_{\perp 2} \rightarrow 0$), $|t(k_z)| \rightarrow t_{\perp 1}$
and 
\begin{equation}
\mbox{Re} [\sigma_{zz}(\omega) ] \rightarrow
 \frac{2 e^2 N_\| t_{\perp 1}^2 d_1^2}
{\hbar^2 d} \, \frac{\tau}{\tau^2(\omega - 2 t_{\perp 1}/\hbar)^2 + 1}.
\end{equation}
The $c$-axis response becomes a Lorentzian centred at $\hbar \omega  = 
2t_{\perp 1}$.  This case has been studied in detail by Gartstein,
Rice and van der Marel.\cite{Gartstein}

\section{Sum Rules}
\label{sum}

In this section we will discuss the partial sum rule for the conductivity
within our model.  The full sum rule
\begin{equation}
\label{30}
\frac{\pi n e^2}{2 m} = \int_0^\infty d\omega \, \sigma(\omega),
\end{equation}
where $n$ is the electron density and $m$ is the bare mass, is well known.
Equation~(\ref{30}) is often used as a means of measuring the electron
density in the cuprate superconductors.

In any practical evaluation of Eq.~(\ref{30}), it is necessary to
impose a cutoff in the frequency integration, and often this cutoff is
taken to be below the onset of interband transitions in the $a$ and
$b$ directions, so that only the Drude-like response is counted.
If we define, therefore, a plasma frequency tensor by
\begin{equation}
\frac{\omega_p^2}{8} = \int_0^\infty d\omega \, 
\sigma_{\mbox{\scriptsize Drude}}(\omega)
\end{equation}
and consider $\Gamma \rightarrow 0$ for simplicity then, from
Eq.~(\ref{17a}), 
\begin{equation}
\frac{\omega_{p\,\mu\nu}^2}{8} = -\frac{\pi e^2}{\Omega} \sum_\pm 
\sum_{\bf k}
\left[ \frac{1}{\hbar^2} \frac{\partial \epsilon_\pm}{\partial k_\mu} 
\frac{\partial \epsilon_\pm}{\partial k_\nu} \right ]
\frac{\partial f(\epsilon_\pm)}{\partial \epsilon_\pm} .
\end{equation}
Integrating by parts in $k_\mu$ gives
\begin{equation}
\frac{\omega_{p\,\mu\nu}^2}{8} = \frac{\pi e^2}{\Omega} \sum_\pm 
\sum_{\bf k}
\left[ \frac{1}{\hbar^2} \frac{\partial^2 \epsilon_\pm}{\partial k_\mu 
 \partial k_\nu} \right ] f(\epsilon_\pm).
\end{equation}
Now, we can define an average effective mass tensor for each of the bands
by
\begin{equation}
\overline{M^{-1}_{\pm\,\mu\nu}} = \frac{1}{n_\pm}\times\frac{2}{\Omega}
\sum_{\bf k}
\left[ \frac{1}{\hbar^2} \frac{\partial^2 \epsilon_\pm}{\partial k_\mu 
 \partial k_\nu} \right ] f(\epsilon_\pm),
\end{equation}
where
\begin{equation}
n_\pm = \frac{2}{\Omega} \sum_{\bf k} f(\epsilon_\pm),
\end{equation}
so that
\begin{equation}
\label{28}
{\omega_{p\,\mu\nu}^2} = {4\pi e^2} \sum_\pm {n_\pm}
\overline{M^{-1}_{\pm\,\mu\nu}}.
\end{equation}
We emphasize here 
that $\overline{M^{-1}_{\pm\,\mu\nu}}$ is an average of the effective
mass tensor {\em over all filled states}, and is not just the effective
mass tensor at the Fermi surface.  As a result of this, 
$\overline{M^{-1}_{\pm\,\mu\nu}}$ depends on the filling of the bands.
For this reason, we suggest that it may actually be difficult to 
determine changes in the electron density with doping in the high $T_c$
materials.

\section{Conclusion and Discussion}
\label{conc}

We have calculated the optical conductivity for a simple layered model of
YBCO$_{7-\delta}$, in which each unit cell consists of a two-dimensional plane
 layer and a one-dimensional chain layer.  The model contains two important 
pieces of physics.  First, we assumed that the layers are coherently coupled
and measured the strength of the coupling with the parameter $t_{\perp}$.
As we mentioned earlier, this is different from the more usual
point of view which
ascribes the broad background and absence of a Drude peak in the $c$-axis
conductivity to incoherent $c$-axis transport.  The second important piece
of physics is the scattering rate $\Gamma$, which was assumed to
vary linearly with $T$. 

In a multilayer system, we found that there are two contributions to
the optical conductivity:  intraband and interband.  For the in-plane
conductivity ($\sigma_{xx}$ and $\sigma_{yy}$) we found that the
intraband conductivity dominates the response, resulting in Drude-like
conductivities.  In fact, we showed that
the intraband terms scale as $(t_\perp)^0$, while the interband terms
scale as $t_\perp^2$ for small $t_\perp$.  On the other hand, we found
that the $c$-axis response is dominated by the interband contribution
(which still scales as $t_\perp^2$) since the intraband contribution
scales as $t_\perp^4$.  For this reason, we found that the $c$-axis
optical conductivity has a non-Drude appearance.

In the plane-chain model, the $\sigma_{zz}$  consists of a broad
featureless response at high temperatures.  As the temperature is lowered,
structure develops in the conductivity.  In some cases, we found that
a pseudogap-like structure appeared.  The broad range of frequencies
spanned by $\sigma_{zz}$ is the result of the plane and chain bands having
different geometries so that transitions between the two bands span a wide
range of energies.   In contrast, 
we showed that a bilayer system in which the two
layers have identical band structures results in a $c$-axis conductivity
which is finite only over a narrow range of frequencies.

The other feature of $\sigma_{zz}$---that structure appears as $T$ is
lowered---is the result of having a temperature dependent scattering rate.
At $T=100$ K, for example, structure in $\sigma_{zz}$ is smeared out over
20 meV, while at $T = 10$ K, structure can be resolved on a scale of
2 meV.

Finally, with this model, we have been able to comment on changes
in $\sigma_{zz}$ with doping over the range YBCO$_{6.7}$ to YBCO$_7$.
We have suggested that in slightly underdoped YBCO, there is a band gap 
between the plane and chain layers, which is reflected in a gap in the
interband conductivity.  Above a certain temperature, however, the band
gap is hidden by the large scattering rate which smears out quasiparticle 
energies by more than the band gap.  As the temperature is reduced, the
quasiparticle energies become better defined and the band gap appears in
the conductivity.  The gap is similar in its appearance to the 
pseudogap observed in YBCO$_{6.7}$.  At optimal doping,
we have suggested that two changes must be made to the model.  The first
is that the strength of the chain-plane coupling must be increased---thus
increasing the importance of the Drude contribution to $\sigma_{zz}$.  The
second is that the chain and plane Fermi surfaces must be made to cross,
eliminating the pseudogap.

We point out that the exact shape of the interband contribution as a 
function of energy was found to be quite sensitive to details of the Fermi 
surfaces involved.  This means that $c$-axis conductivity measurements 
could, in principle, be used to get information on the energy bands 
as well as on the filling factors for chains and planes and on their changes 
with oxygen doping.  

While the sensitivity of the frequency dependence of $\sigma_c(\omega)$ to 
band structure could, in principle, be used to get spectroscopic information
on electronic structure, it should be emphasized that the band structure 
used in our work is grossly oversimplified and so  some of our detailed 
predictions cannot be applied directly to YBa$_2$Cu$_3$O$_x$.  The 
qualitative features obtained and emphasized in this conclusion are, however, 
expected to be robust and remain in more complex models.  Such calculations 
will need to employ more realistic band structures and, perhaps more 
importantly, consider the issue of intercell coupling which could be 
incoherent and quite different from the chain-plane hopping.  Nevertheless, 
our model does exhibit many of the features observed in experiments on 
YBCO at optimum doping as well as underdoped and overdoped cases.

\section*{Acknowledgments}
This work was supported by a Natural Sciences and Engineering
Research Council of Canada (NSERC) grant, and by the Canadian Institute for 
Advanced Research (CIAR).

\begin{figure}
\begin{center}
\leavevmode
\epsfxsize 0.9\columnwidth
\caption{Fermi surface for the chain-plane model.  The Fermi surface is
shown for $k_z = 0$ (solid curve) and $k_z = \pi/d$ (dashed curve).  When
$k_z  = \pi/d$, the chain-plane coupling $t({\bf k})$ vanishes and the Fermi
surface is that of the isolated chain and plane layers.  The dotted curves
are lines of the constant energy difference $\omega = \epsilon_+ - \epsilon_-$.
The band structure
parameters chosen for this case are
$\{t_1,t_2,\mu_1,\mu_2,t_\perp\} = \{70,100,-65,-175,20\} \mbox{ meV}$ and $B = 0.45$. }
\label{f1}
\end{center}
\end{figure}

\begin{figure}
\begin{center}
\leavevmode
\epsfxsize \columnwidth
\caption{Normal state optical conductivity in the (a) $x$ (b) $y$ and (c) $z$
directions for the plane-chain model.  The conductivity is shown for 
temperatures 200 K (solid curve), 100 K (dot-dashed curve), 10 K (dashed
curve) and 0 K (dotted curve).  The scattering rate is $1/\tau = 20$ meV
at 100 K and it scales linearly with temperature.  The conductivity in
the $z$-direction is dominated by interband processes and has a non-Drude
appearance, while the conductivity in the $x$ and $y$ directions has
a predominantly Drude-like behaviour, although there is a small interband 
contribution to $\sigma_{yy}$.  The band structure parameters are the same 
as in Fig.~\protect\ref{f1}, except that $t_\perp = 5$ meV.}
\label{f2}
\end{center}
\end{figure}

\begin{figure}
\begin{center}
\leavevmode
\epsfxsize \columnwidth
\caption{The normal state conductivity is shown for the (a) $x$, (b) $y$
and (c) $z$ directions at 100~K (solid curve) and 0~K (dotted curve).
The band structure differs from that of Fig.~\protect\ref{f2} by the
magnitude of the chain-plane coupling, which is $t_\perp = 10$ meV here.}
\label{f3}
\end{center}
\end{figure}

\begin{figure}
\begin{center}
\leavevmode
\epsfxsize \columnwidth
\caption{The normal state conductivity is shown, as in Fig.~\protect\ref{f3},
but with $t_\perp = 20$ meV.  The conductivity in the $z$ direction
has a significant Drude part, while the conductivity in the $y$ direction
has a significant interband part.}
\label{f4}
\end{center}
\end{figure}

\begin{figure}
\begin{center}
\leavevmode
\epsfxsize \columnwidth
\caption{The Fermi surface is shown for the chain-plane model for a
case where the chain and plane Fermi surfaces (dashed curves) cross.
This case is qualitatively different from the case shown in Fig.~\protect%
\ref{f1} because there is no band gap.  As a result, a pseudogap is
not expected in the $c$-axis optical conductivity.  The band structure
parameters chosen for this case are $\{t_1,t_2,\mu_1,\mu_2,t_\perp\}
= \{70,100,-65,-130,20\} \mbox{ meV}$ and $B = 0.45$. }
\label{f5}
\end{center}
\end{figure}

\begin{figure}
\begin{center}
\leavevmode
\epsfxsize \columnwidth
\caption{The optical conductivity is shown along the (a) $x$, (b) $y$ and
(c) $z$ directions at $T=100$ K and at $T=0$ K.  The $T=0$ K optical
conductivity is entirely due to interband transitions at finite frequencies.
In $\sigma_{xx}$ there is only a very small interband contribution,
while in $\sigma_{yy}$ and $\sigma_{zz}$, the interband contributions
are substantial.  At $T=100$ K, however, it is difficult to distinguish
the interband contribution from the Drude contribution because of the large
scattering rate.  There is also a large Drude contribution to $\sigma_{zz}$
and, and there is no pseudogap, although the interband conductivity
still falls to zero linearly with $\omega$.
The band parameters are the same as in Fig.~\protect%
\ref{f5}, except that $t_\perp = 10$ meV.}
\label{f6}
\end{center}
\end{figure}

\begin{figure}
\begin{center}
\leavevmode
\epsfxsize \columnwidth
\caption{The optical conductivity is shown, as in Fig.~\protect\ref{f6},
but with $t_\perp = 20$ meV.}
\label{f7}
\end{center}
\end{figure}

\begin{figure}
\begin{center}
\leavevmode
\epsfxsize 0.9\columnwidth
\caption{The Fermi surface is shown for the bilayer system.  In this model,
there are two plane layers per unit cell.  The dispersions in the layers
are the same, but they are spaced distances $d_1$ and $d_2$ apart in
alternating fashion.  The Fermi surfaces are shown for $k_z = 0$
(solid line) and $k_z = \pi/d$ (dashed line).  The minimum and 
maximum coupling strengths are $2|t_{\perp 1} - t_{\perp 2}|$ and 
$2|t_{\perp 1} + t_{\perp 2}|$ respectively.  The model parameters are
$\{t_1,\mu_1,t_{\perp 1},t_{\perp 2}\} = \{70,-65,20,10\} \mbox{ meV}$,
$d_1 = 0.3 d$ and $B= 0.45$.}
\label{f9}
\end{center}
\end{figure}

\begin{figure}
\begin{center}
\leavevmode
\epsfxsize 0.9\columnwidth
\caption{The optical conductivity is shown for the bilayer system 
(a) in the $a$ and $b$-directions and (b) in the $c$-direction.  
The conductivity is shown at $T=100$ K (solid curve) and at 
$T=0$ K (dotted curve).  The in-plane conductivity has a Drude shape,
while the $c$-axis conductivity has a large interband contribution.
At $T=0$ K, the in-plane conductivity vanishes at finite frequency.
Along the $c$-axis, however, the interband contribution remains for
frequencies $2|t_{\perp 1}-t_{\perp 2}| \leq \hbar \omega
2|t_{\perp 1}+t_{\perp 2}|$.  The model parameters are the same as 
those in Fig.\ \protect\ref{f9}.}
\label{f10}
\end{center}
\end{figure}

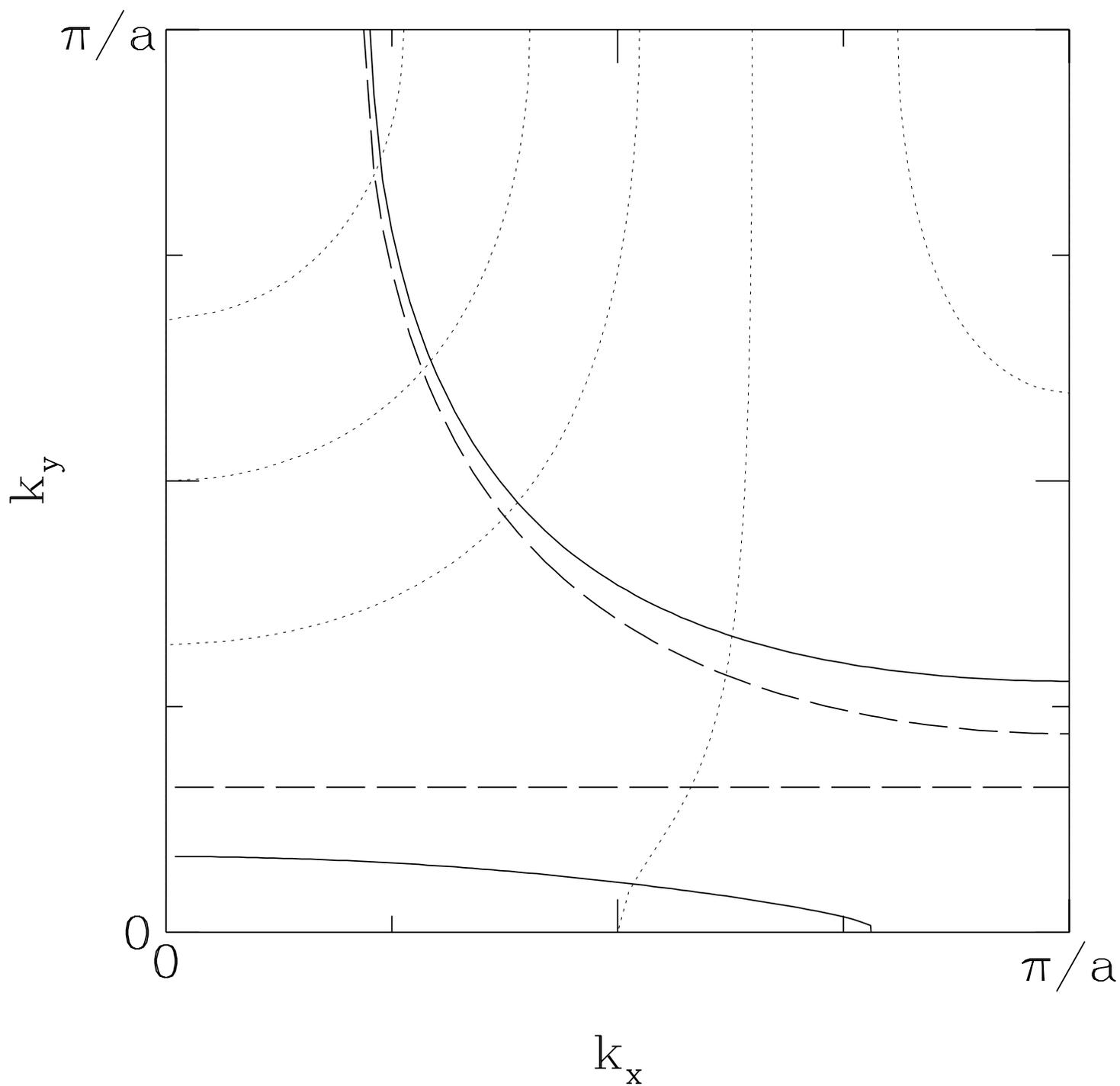
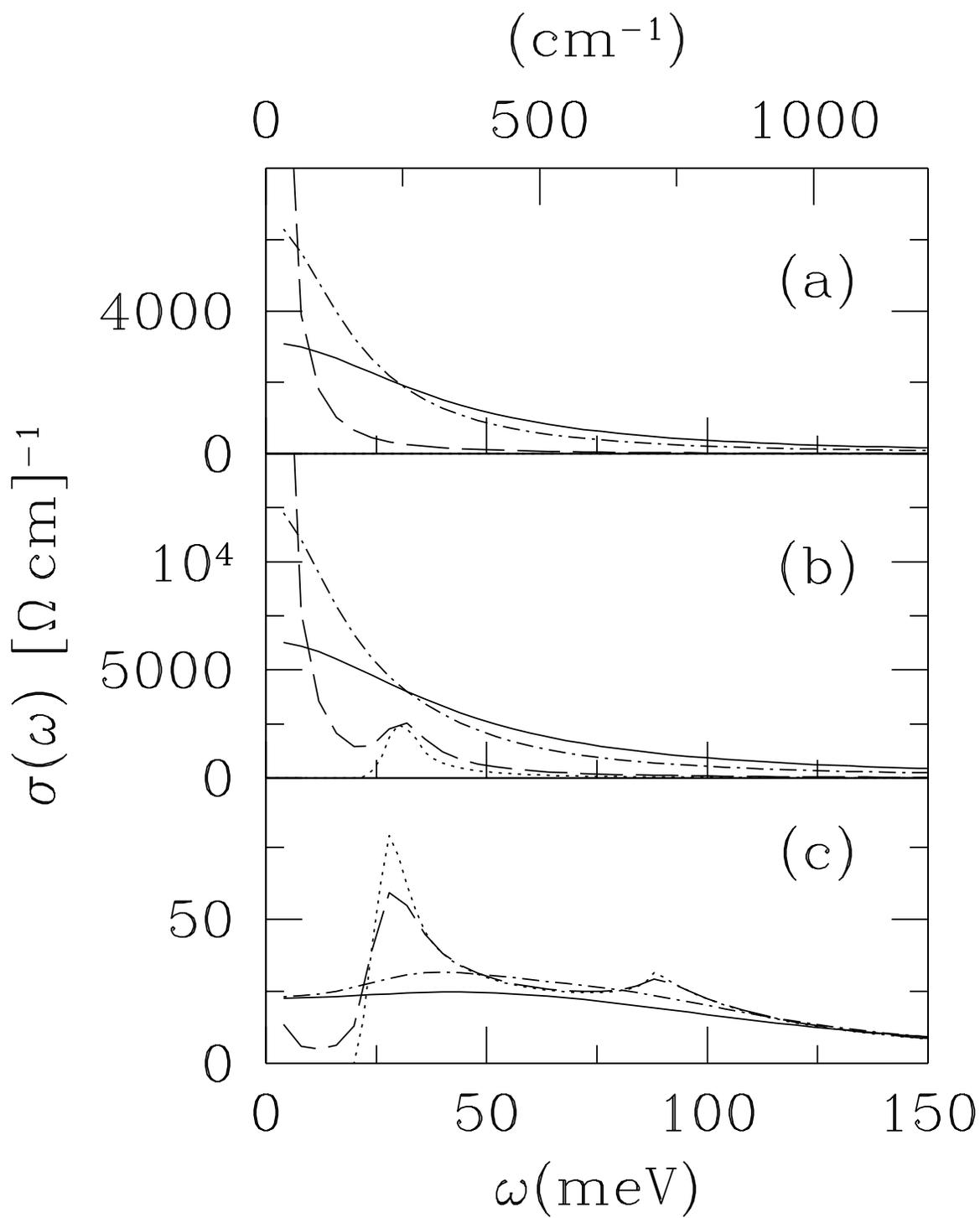
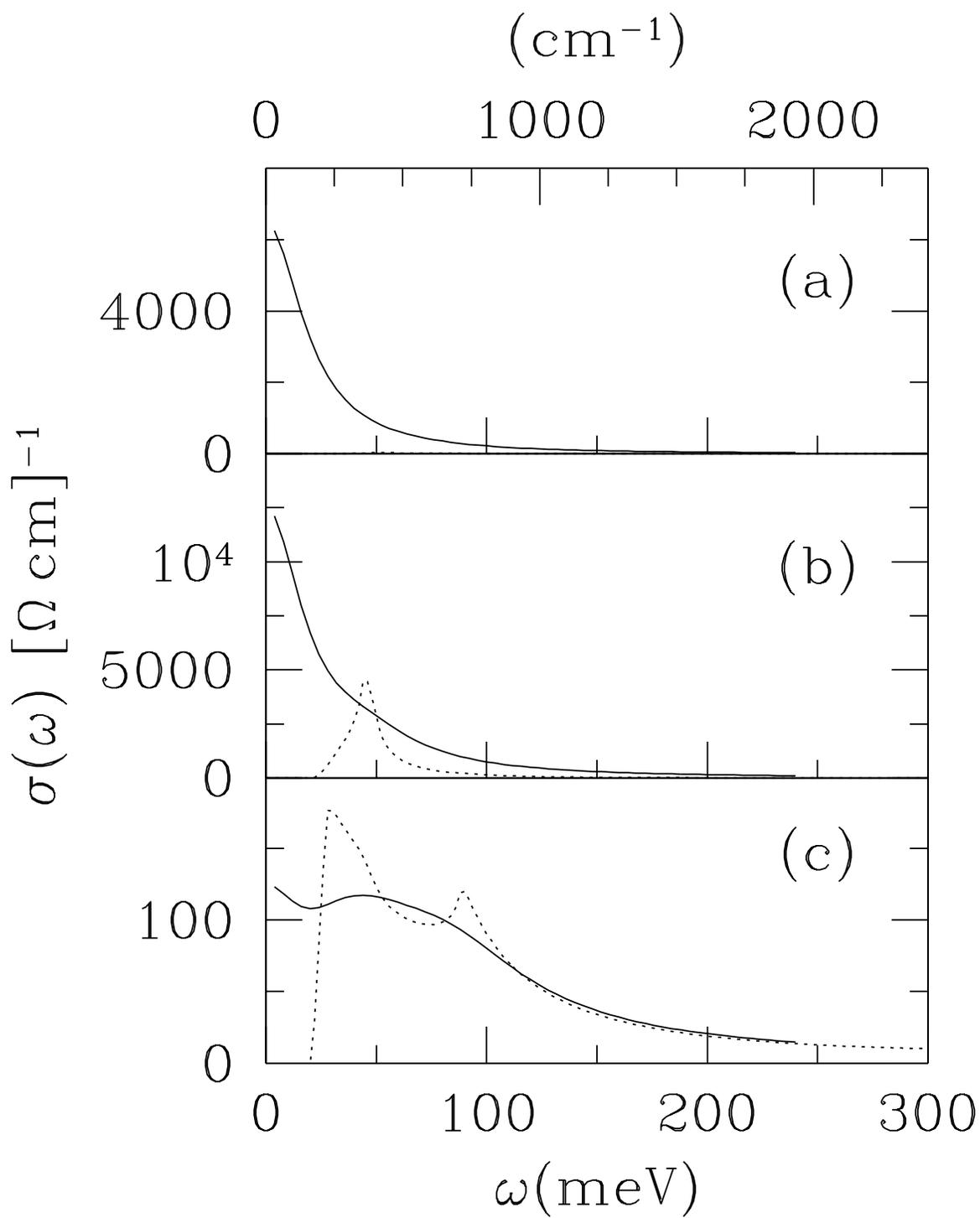
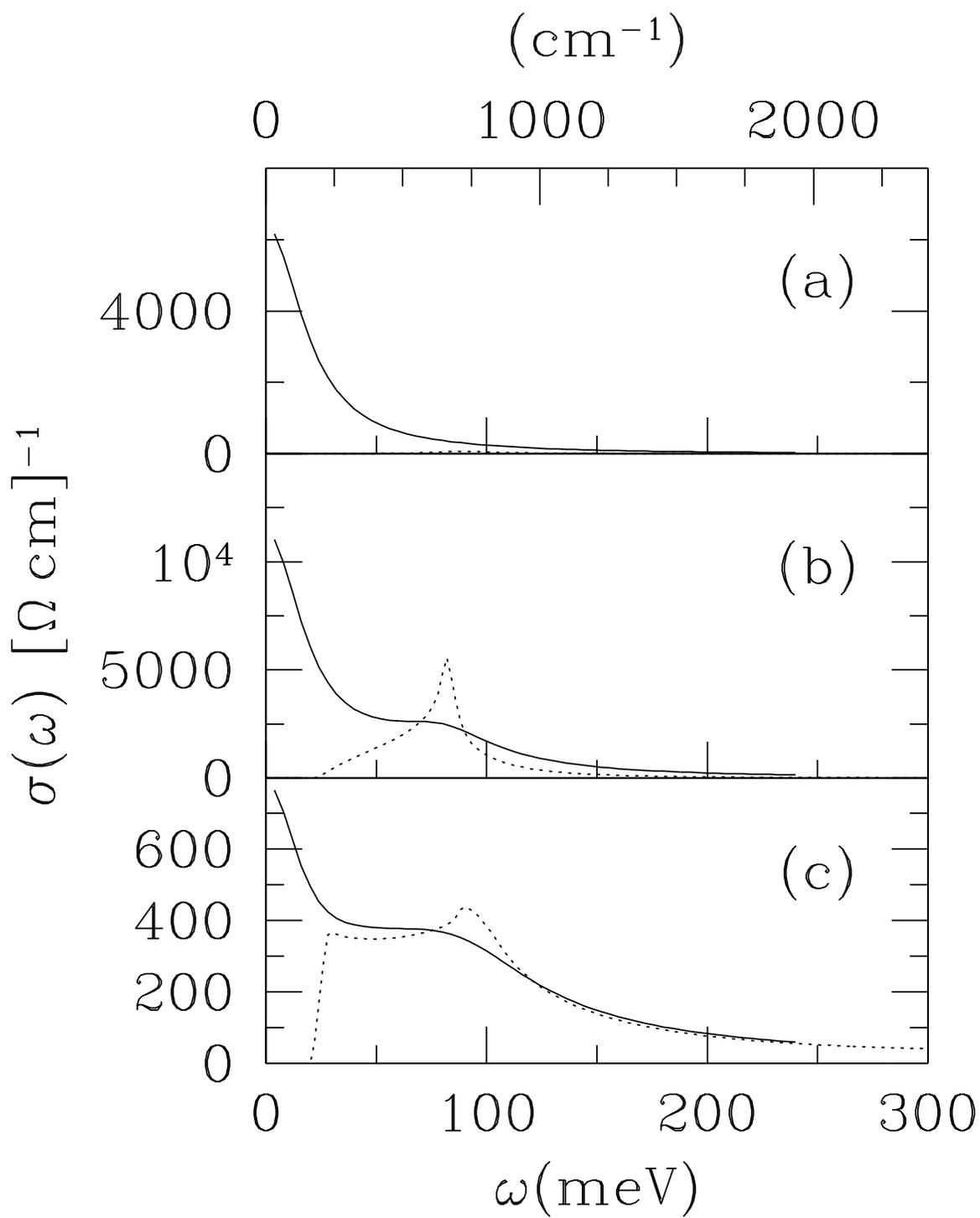
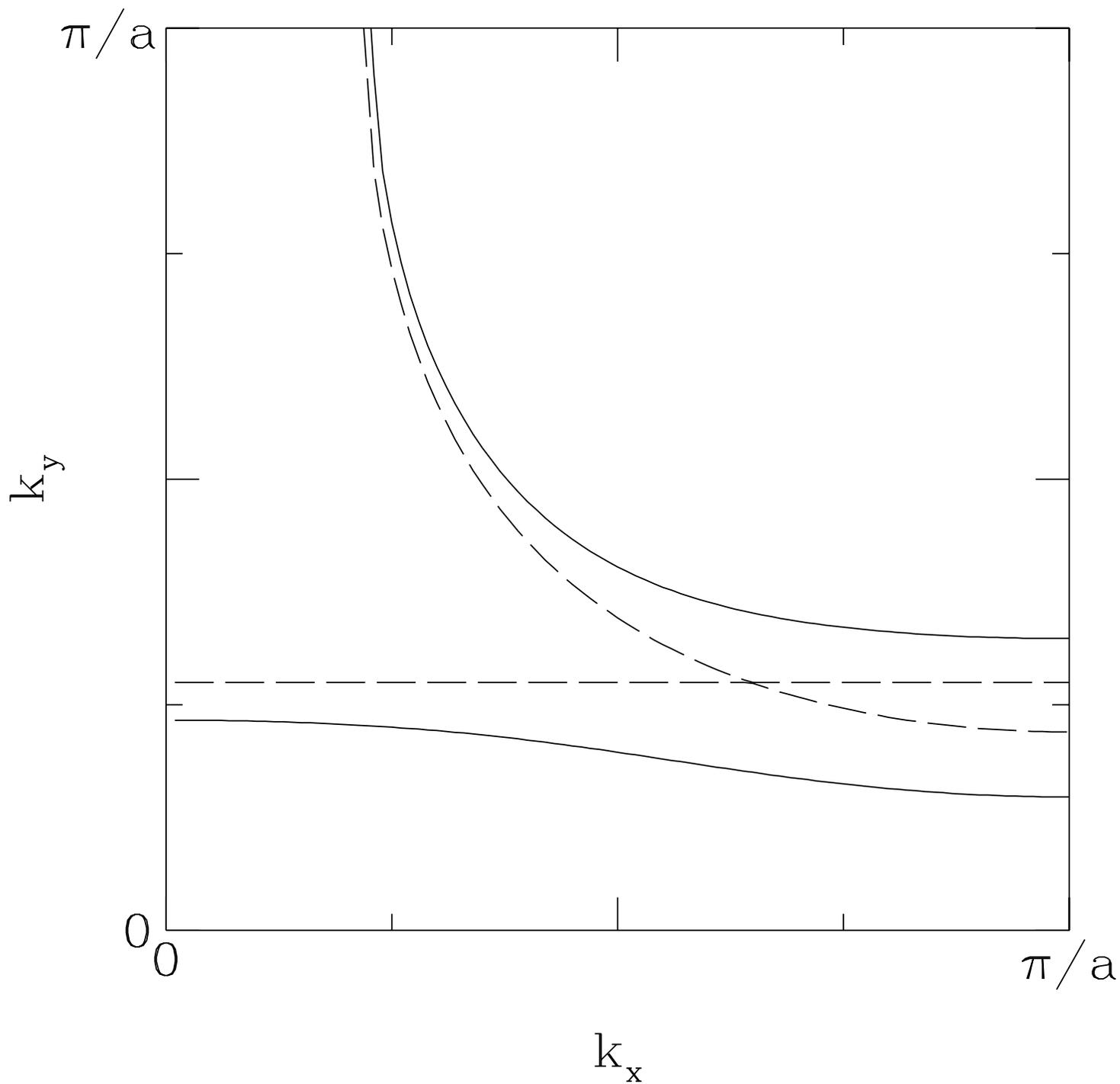
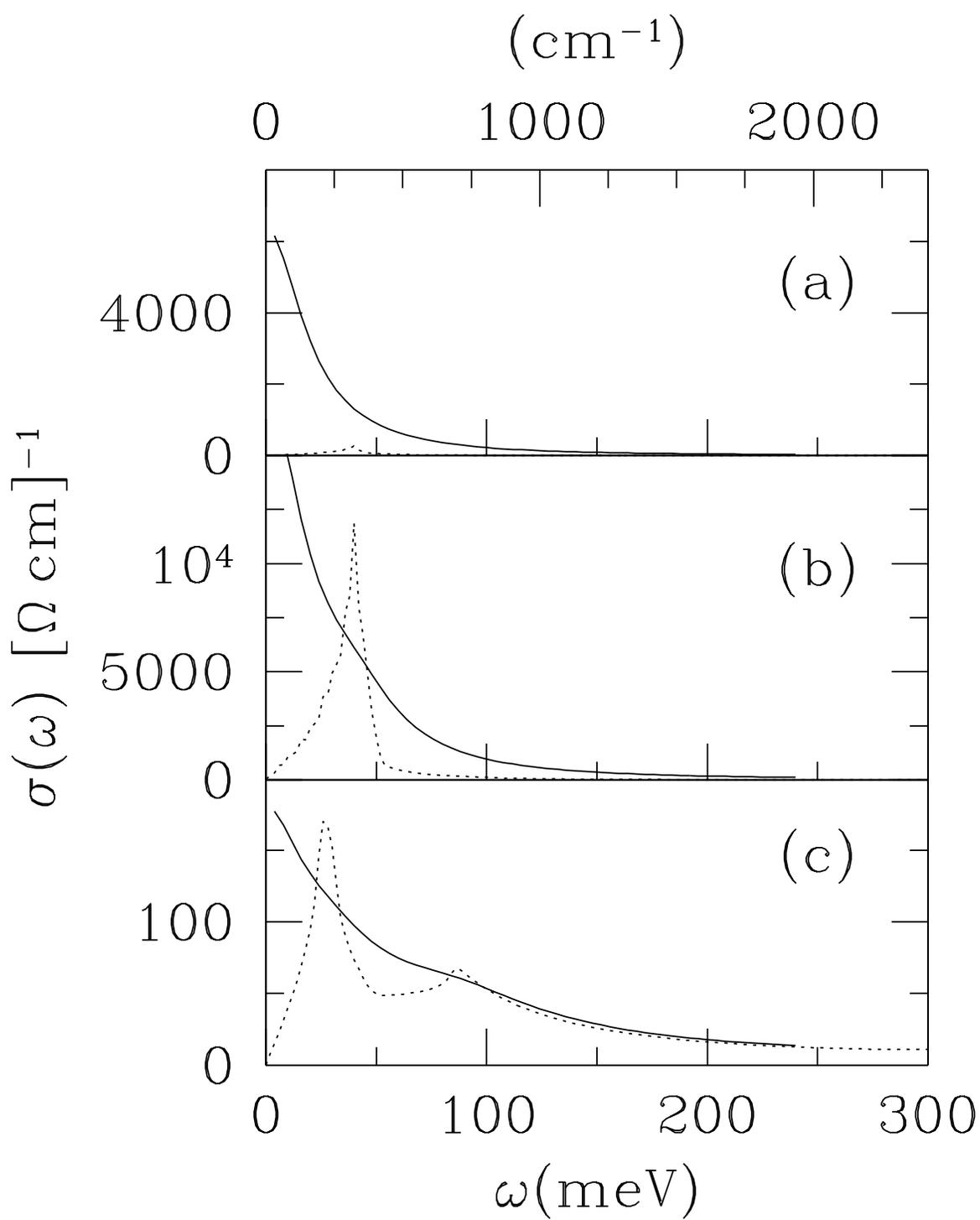
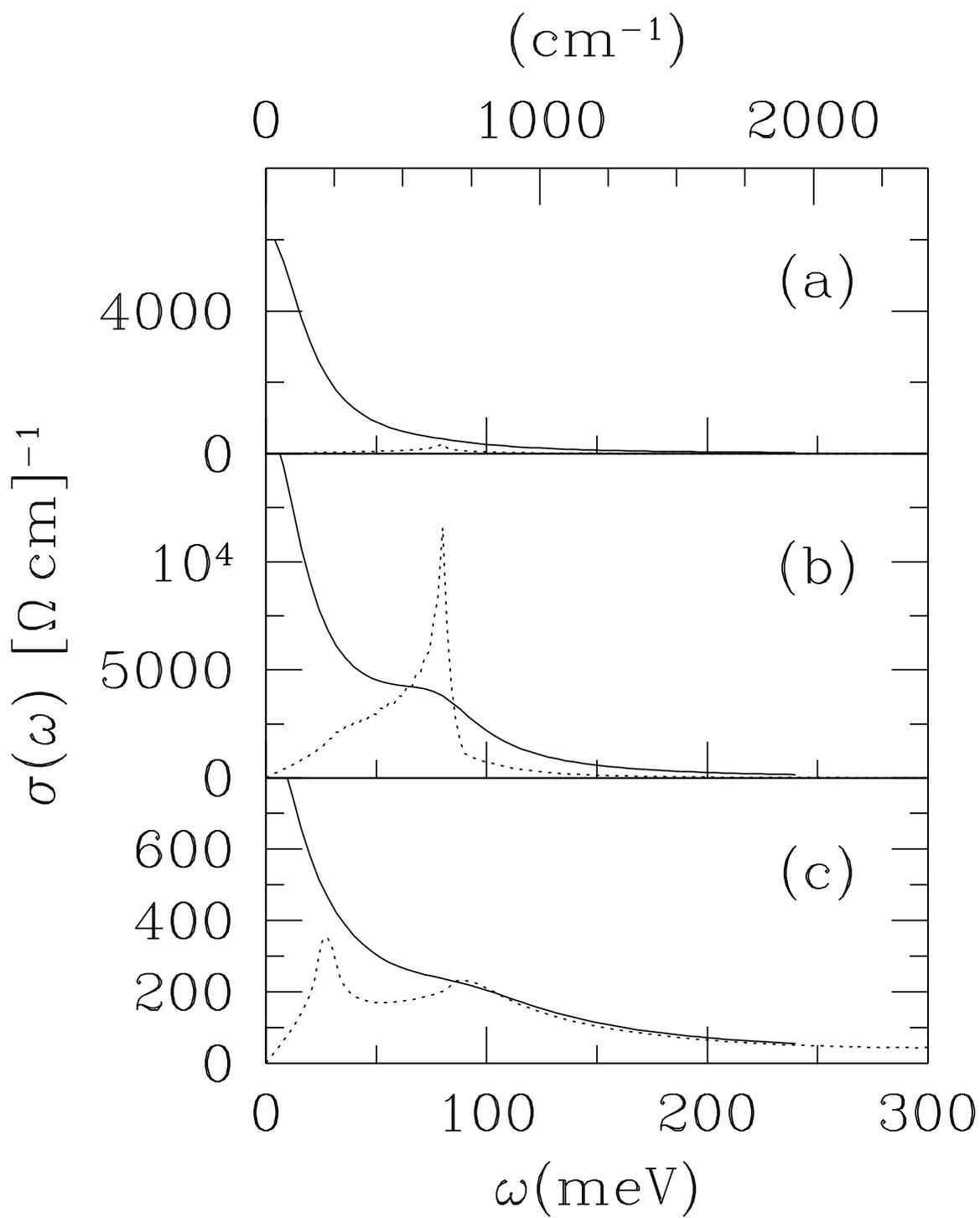
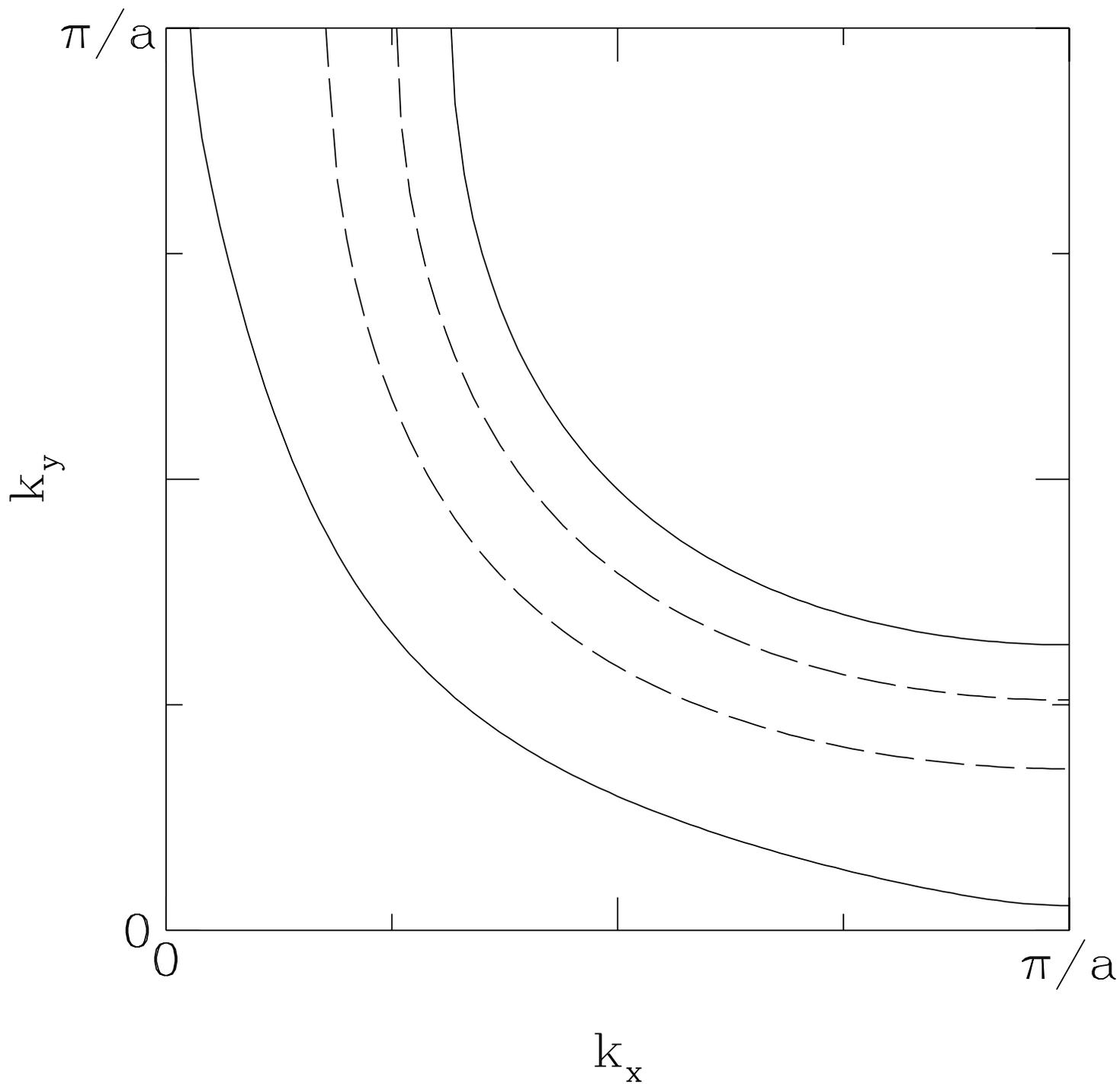
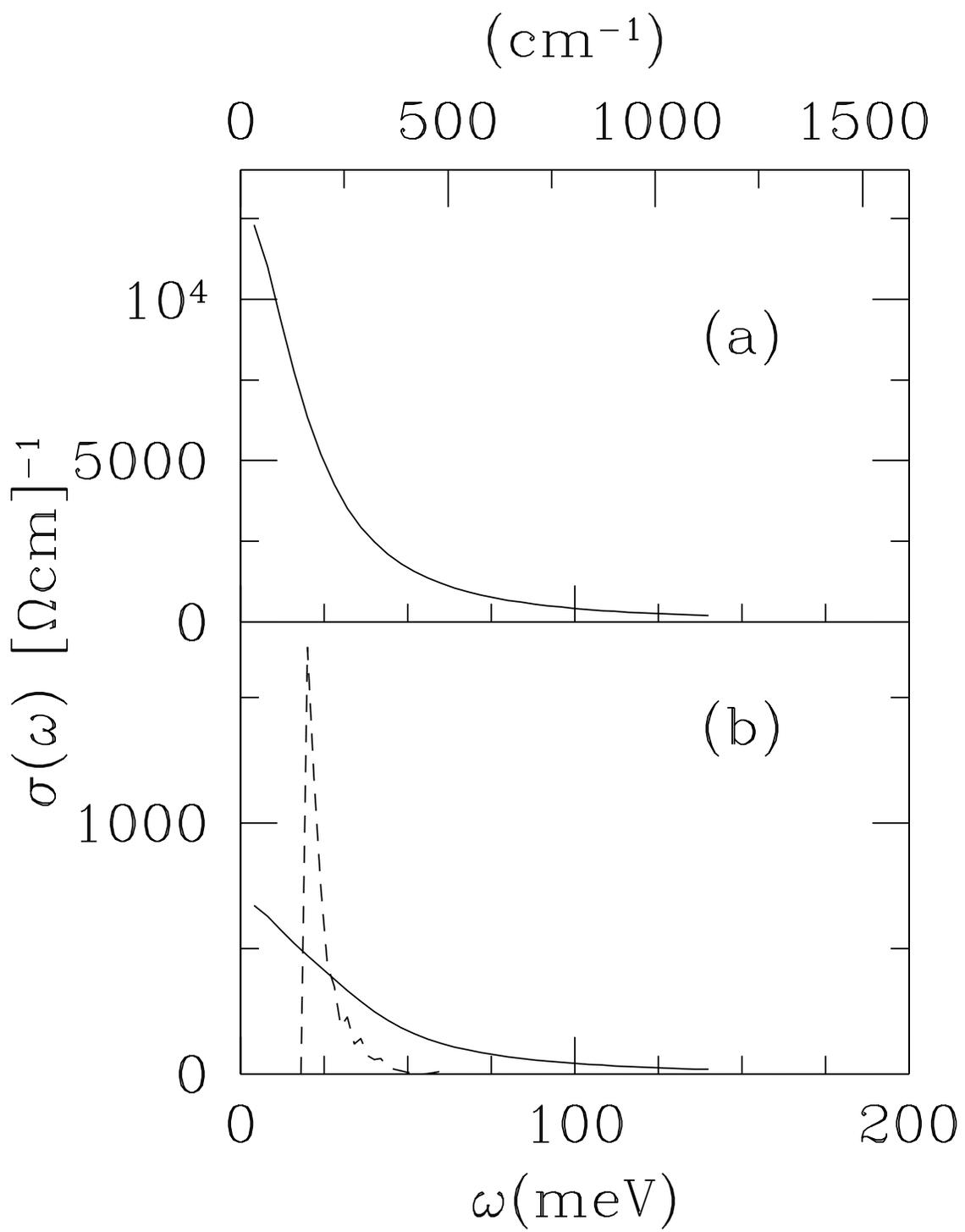
\end{document}